\documentclass[sigconf]{acmart}
\usepackage[capitalize]{cleveref}
\usepackage{algorithm}
\usepackage{algorithmic}
\usepackage{changepage}

\usepackage{listings}
\AtBeginDocument{%
  }

\setcopyright{acmlicensed}
\copyrightyear{2026}
\acmYear{2026}
\setcopyright{cc}
\setcctype{by}
\acmConference[DAC '26]{63rd ACM/IEEE Design Automation Conference}{July 26--29, 2026}{Long Beach, CA, USA}
\acmBooktitle{63rd ACM/IEEE Design Automation Conference (DAC '26), July 26--29, 2026, Long Beach, CA, USA}
\acmDOI{10.1145/3770743.3803992}
\acmISBN{979-8-4007-2254-7/2026/07}

\begin{document}

\settopmatter{printacmref=false}

\title{CUTEv2: Unified and Configurable Matrix Extension for Diverse CPU Architectures with Minimal Design Overhead}


\makeatletter
\def\@authorfont{\fontsize{11.4pt}{13.2pt}\selectfont}
\def\@authornotemark{}
\makeatother

\author{Jinpeng Ye$^{1,2}$, Chongxi Wang$^{1,2,*}$, Wenqing Li$^{1,2}$, Bin Yuan$^{1,2}$, Shiyi Wang$^{1,2}$, Fenglu Zhang$^{1,2}$, Junyu Yue$^{1,2}$,
Jianan Xie$^{1,2}$, Yunhao Ye$^{1,2}$, Haoyu Deng$^{1,2}$, Yingkun Zhou$^{1,2}$, Xin Cheng$^{1,2}$, Fuxin Zhang$^{1,2}$, Jian Wang$^{1,2}$}
\affiliation{%
  \institution{$^1$State Key Lab of Processors, Institute of Computing Technology, Chinese Academy of Sciences, Beijing, China}
}
\affiliation{%
  \institution{$^2$University of Chinese Academy of Sciences, Beijing, China}
}
\authornote{Corresponding author is Chongxi Wang: wangchongxi@ict.ac.cn.}

\renewcommand{\shortauthors}{Jinpeng Ye et al.}




\begin{abstract}

Matrix extensions have emerged as an essential feature in modern CPUs to address the surging demands of AI workloads. However, existing designs often incur substantial hardware and software design overhead. Tight coupling with the CPU pipeline complicates integration across diverse CPUs, while fine-grained synchronous instructions hinder the development of high-performance kernels.

This paper proposes a unified and configurable CPU matrix extension architecture. By decoupling matrix units from the CPU pipeline, the design enables low-overhead integration while maintaining close coordination with existing compute and memory resources. The configurable matrix unit supports mixed-precision operations and adapts to diverse compute demands and memory bandwidth constraints. An asynchronous matrix multiplication abstraction with flexible granularity conceals hardware details, simplifies matrix-vector overlap, and supports a unified software stack.

The architecture is integrated into four open-source CPU RTL platforms and evaluated on representative AI models. Matrix unit utilization under GEMM workloads exceeds 90\% across all platforms. When configured with compute throughput and memory bandwidth comparable to Intel AMX, our design achieves speedups of 1.57×, 1.57×, and 2.31× on ResNet, BERT, and Llama3, with over 30\% of the gains attributed to overlapped matrix-vector execution. A 4 TOPS@2GHz matrix unit occupies only 0.53 mm\textsuperscript{2} in 14nm CMOS. These results demonstrate strong cross-platform adaptability and effective hardware–software co-optimization, offering a practical matrix extension for the open-source community.$^1$ \footnote{$^1$Code available at \url{https://github.com/OpenCUTE/CUTE}.}

\end{abstract}

\maketitle

\section{Introduction}

The rapid advancement of AI technologies has driven widespread deployment across domains such as natural language processing, computer vision, multimodal generation, and embodied intelligence, spanning platforms from edge devices to data centers\cite{mlperf,Mixed-precisionMatrixMultiplication}. As a foundational component of general-purpose computing systems, CPUs continue to play a critical role in executing diverse AI tasks\cite{ktransformers}. To address the surging demand for matrix-intensive computation, CPU vendors have integrated matrix extensions including Intel AMX\cite{intel_amx}, Arm SME\cite{arm_sme}, and IBM MMA\cite{ibm_mma} into products. The RISC-V community has also proposed several matrix extension standards such as IME\cite{riscv_ime}, VME\cite{riscv_vme}, and AME\cite{riscv_ame}. These extensions effectively leverage existing compute and memory resources, particularly vector units and cache subsystems, to significantly enhance AI performance with modest area and power overheads.

However, existing CPU matrix extensions pose challenges for hardware integration and software programmability\cite{asplos2026_Configurationwall}. Architecturally, matrix units are often tightly coupled with the CPU pipeline, entailing close interaction with vector register files or load-store units. Fine-grained synchronous matrix instructions introduce substantial structural and data hazards, increasing integration complexity and verification effort. This coupling limits portability across microarchitectures. On the software side, AI workloads typically combine matrix multiplication with a large number of element-wise operations, requiring close coordination between matrix and vector units. Expressing such fine-grained interleaving within a single instruction stream places significant burden on both programmers and compilers, complicating kernel design and scheduling.

To address these challenges, this paper proposes a unified and configurable CPU matrix extension architecture designed for agile integration and efficient execution across platforms. The matrix unit is carefully decoupled from the CPU pipeline to avoid intrusive modifications to register files and memory paths, reducing co-design and verification complexity. To accommodate diverse compute and bandwidth constraints, it supports flexible microarchitectural configurations guided by a compute-bandwidth constraint model, and can be scaled from 0.5 to 32 TOPS to guarantee resource utilization and platform adaptability. The matrix unit supports FP8/INT8/FP16/BF16/TF32 mixed-precision computing to meet varying accuracy and performance requirements. As for ISA and programming model, only an asynchronous matrix multiplication and a synchronization primitive are defined, forming a minimal and unified interface that supports flexible granularity. This abstraction hides hardware-specific details and simplifies the programming of overlapped matrix-vector execution, improving programmability and enabling a portable software stack.

We integrate and validate the proposed architecture on four open-source CPU RTL platforms: Rocket\cite{rocket}, Shuttle\cite{shuttle}, BOOM\cite{boom}, and XiangShan-Kunminghu\cite{kunminghu}. We further evaluate GEMM and AI inference performance against Intel AMX, Arm SME, and IBM MMA. The matrix unit achieves over 90\% utilization on GEMM workloads across all integrated platforms. On the Shuttle CPU with a 512-bit Saturn~\cite{saturn} vector unit, a 4 TOPS@8-bit Matrix Unit and 48 GB/s memory bandwidth, the design delivers 1.57×, 1.57×, and 2.31× speedups on ResNet\cite{resnet50}, BERT\cite{bert}, and Llama3\cite{grattafiori2024llama} inference compared to Xeon 8580. Overlapped matrix–vector execution contributes 66.7\%, 50.9\%, and 33.6\% of the performance gain on three workloads.The design also outperforms IBM S1022 MMA (8.87×, 3.33×, 3.08×) and Apple M4 SME (5.04×, 2.11×, 3.16×). A 4 TOPS@2GHz matrix unit occupies only 0.53 mm\textsuperscript{2} in 14nm CMOS.

The key contributions of this paper are as follows:
\begin{itemize}
\item Proposing a unified and configurable CPU matrix extension architecture that enables agile cross-platform integration and efficient execution.
\item Presenting a co-designed hardware–software solution that delivers substantial performance gains on representative AI models over commercial CPU matrix extensions.
\item Integrating the proposed matrix extension into four open-source CPU RTL platforms, with all RTL implementations and high-performance kernels fully open-sourced.
\end{itemize}

\vspace{-10pt}
\section{Background}
\subsection{AI Workloads}

AI workloads have been widely deployed across diverse computing platforms, from edge devices to data centers. Typical models consist of layers with heterogeneous characteristics. Compute-intensive layers (e.g., linear, attention, convolution) are dominated by matrix multiplications, while element-wise operations (e.g., activation, (de)quantization, normalization) are generally memory intensive.

\begin{figure}[t]
\centering
\vspace{-10pt}
\includegraphics[width=0.9\linewidth]{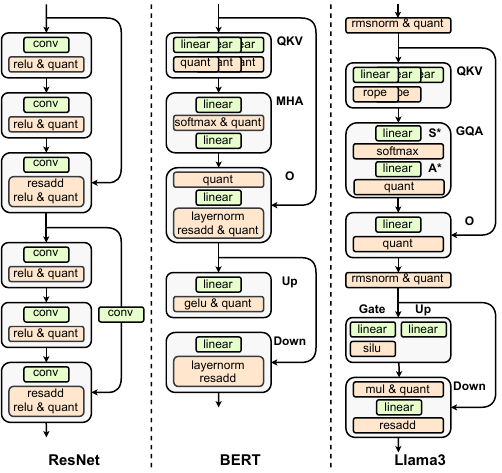}
\vspace{-10pt}
\caption{AI Model Architectures and Kernel Fusion Patterns.}
\label{fig:ai_workload}
\vspace{-15pt}
\end{figure}

Figure \ref{fig:ai_workload} illustrates the architecture and kernel fusion strategies of three representative models - ResNet, BERT, and Llama3. Kernel fusion\cite{xFormers2022,flashattention,flashattention2} enhances overall performance by exploiting data locality and fusing operators into tiled pipelines, thereby reducing memory traffic and improving resource utilization.

\vspace{-8pt}
\subsection{Related Work}\label{sec:relatedwork}

\begin{figure}[t!]
\centering
\includegraphics[width=1.1\linewidth]{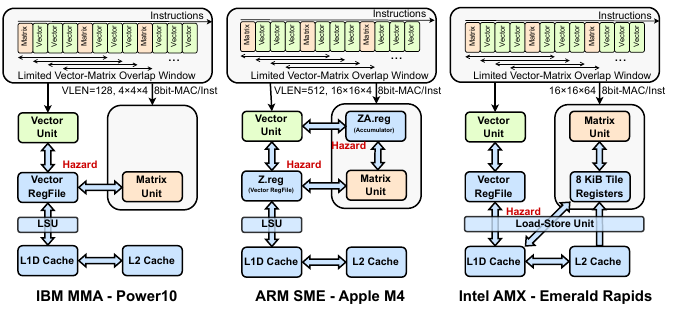}
\vspace{-20pt}
\caption{Existing CPU Matrix Extensions.}
\label{fig:related_work}
\vspace{-12pt}
\end{figure}

In recent years, CPU vendors - including Intel, Arm, and IBM - as well as the RISC-V community have introduced matrix extensions to enhance AI capabilities on general-purpose processors. These extensions typically employ fine-grained synchronous matrix instructions and differ in register architectures, execution models, and integration strategies, as illustrated in Figure \ref{fig:related_work}.

IBM MMA and RISC-V IME reuse the data path of the vector unit, repurposing part of the vector register file as accumulators and restricting matrix operations to vector registers. As a result, register size and bandwidth limit the granularity and throughput of matrix operations. IBM Power10 S1022 delivers a per-core INT8 peak of 2 TOPS at 4 GHz.

Arm SME and RISC-V VME introduce dedicated accumulator registers, enabling larger matrix units decoupled from vector register organization. Apple M4 implements SME on both performance and efficiency cores, with the performance core achieving a per-core INT8 peak of 4 TOPS.

Intel AMX and RISC-V AME decouple the matrix unit from the vector pipeline by introducing tile registers and dedicated load–store paths, enabling larger matrix operations. Sapphire Rapids and Emerald Rapids deliver a per-core INT8 peak of 2 TOPS/GHz, with the Xeon 8580 reaching 4.6 TOPS at 2.3 GHz under TDP limits.

The academic community has also explored CPU matrix units in both configurable forms\cite{gemmini, aspdac2025_me} and fixed-size implementations\cite{hpca2020_me, jssc2023_me, jssc2024_me, RiscvTPU}. However, these designs typically lack close cooperation with existing CPU vector units and offer limited scalability and precision support. In comparison, this work introduces an adaptable and configurable CPU matrix extension, providing the open-source community with a practical implementation.

\vspace{-8pt}
\subsection{Motivation}

Although existing CPU matrix extensions have achieved significant throughput improvements, key challenges in hardware integration and software programmability still restrict their broad applicability and efficient execution across diverse CPU and workloads.

For hardware integration, most existing designs tightly couple matrix units with the CPU pipeline, necessitating invasive modifications across instruction decode, dispatch, and execution stages. Matrix units often interact with vector register files or load-store units, creating complex control paths and redundant state management, increasing integration and verification cost. In Arm SME, IBM MMA, and RISC-V IME/VME, matrix and vector instructions contend for vector registers, resulting in structural and data hazards. Although Intel AMX and RISC-V AME introduce independent tile registers and enable direct access to L1D or L2 cache, they remain constrained by synchronous semantics, which require modifications to ensure correctness between matrix and scalar/vector memory operations either by stalling potentially conflicting instructions or by resolving address conflicts within the load-store unit.

For software programmability, most CPU matrix extensions adopt fine-grained synchronous matrix instructions, forcing programmers to manually orchestrate scheduling between matrix and memory operations. Moreover, in AI workloads where matrix and element-wise kernels often require overlapped execution, programming complexity increases significantly. Developers must express memory, matrix, and vector tasks within a single instruction stream and keep all functional units busy within a limited instruction window. Furthermore, disparities among matrix-extension ISAs and software interfaces hinder unified abstractions, requiring developers to perform microarchitecture-specific optimizations and exacerbating software fragmentation.

This paper proposes a unified and configurable CPU matrix-extension architecture, designed to enable agile cross-platform integration and efficient execution with minimal design overhead.

\vspace{-6pt}
\section{Architecture Overview}

This paper proposes a unified and configurable CPU matrix extension, based on three key design principles:
(1) Structural decoupling between the matrix unit and the CPU pipeline, without intrusive modifications to the decoder, instruction issue logic, register file, or load–store units;
(2) Configurable microarchitectural parameters within the matrix unit to accommodate diverse computational requirements and memory system constraints across platforms;
(3) An abstraction for asynchronous matrix-multiplication instructions that enables fusion scheduling with flexible granularity, which simplifies the programming model and improves execution efficiency.

\begin{table}[htbp]
\vspace{-10pt}
\caption{Interface Registers.}
\label{tab:cute_interface}
\vspace{-10pt}
\centering
\setlength{\tabcolsep}{2pt}
\begin{tabular}{lll}
\toprule
\textbf{Field} & \textbf{Type} & \textbf{Description} \\
\midrule
\{M,N,K\}             & uint32  & Matrix Size                        \\
Base \{A,B,Bias,C\}   & uint64  & Memory Base Address                \\
Stride \{A,B,Bias,C\} & uint32  & Memory Stride                      \\
DataType              & enum    & Data Precision                     \\
BiasType              & enum    & Bias Type (Zero, Row-Repeat, Full) \\
Transpose             & bool    & Result Transpose Flag              \\
Status                & uint32  & Async Operation Status             \\
\bottomrule
\end{tabular}
\vspace{-8pt}
\end{table}

\begin{figure}[t!]
\centering
\includegraphics[width=0.87\linewidth]{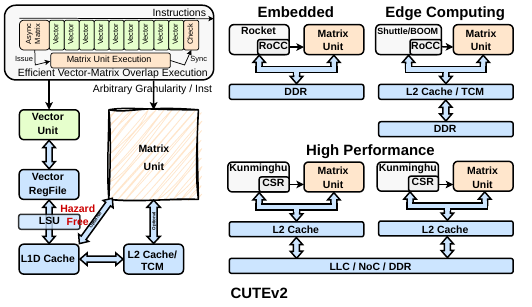}
\vspace{-12pt}
\caption{Architecture Overview.}
\label{fig:Hardware_overview}
\vspace{-15pt}
\end{figure}

Figure \ref{fig:Hardware_overview} illustrates the hardware architecture. The matrix unit is decoupled from the CPU pipeline and driven by asynchronous matrix multiplication instructions. Depending on ISA and microarchitectural support, the CPU dispatches these instructions via a RoCC-like or CSR-based interface, with registers defined in \cref{tab:cute_interface}. The matrix unit connects to cache or memory independently of the CPU’s load–store unit through a platform-adaptable interconnect. This design reduces integration complexity, supports rapid deployment across CPUs, and allows flexible microarchitectural configuration for embedded, edge, and high-performance platforms.

\begin{lstlisting}[caption=Programming Example., label=code:code]
asyncMatMul(TILE_M, TILE_N, K, ...);     // tile 0
for (i=1; i<(M/TILE_M)*(N/TILE_N); i++) {
    asyncMatMul(TILE_M, TILE_N, K, ...); // tile i
    checkMatmul(); // wait tile i-1
    ...            // tile i-1 epilogue
}
checkMatmul();     // wait last tile
...                // last tile epilogue
\end{lstlisting}
\vspace{-5pt}

The asynchronous matrix-multiplication abstraction substantially simplifies programming complexity for matrix extensions and enables efficient fusion of matrix–vector operations. Listing~\ref{code:code} shows a fused kernel example performing matrix multiplication followed by element-wise epilogue computation. The asyncMatMul macro dispatches a task per tile, with tile size determined by shared storage capacity between CPU and matrix unit. During asynchronous execution, the CPU issue window can be fully utilized by the vector unit to compute epilogue operations. Before vector computation proceeds, the checkMatmul instruction ensures the matrix multiplication of the corresponding tile is complete, thereby handling data dependencies correctly.

\section{Design}

\subsection{Matrix Unit Microarchitecture}

\begin{figure}[htbp]
\centering
\vspace{-10pt}
\includegraphics[width=0.9\linewidth]{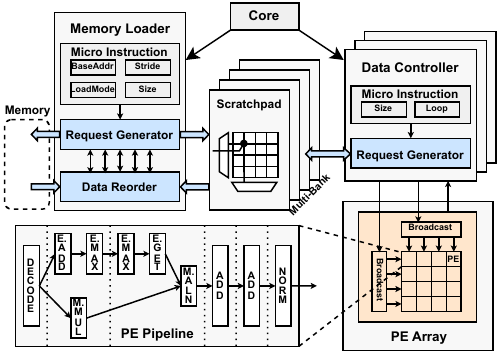}
\vspace{-10pt}
\caption{Matrix Unit Microarchitecture.}
\label{fig:MicroArchitecture}
\vspace{-10pt}
\end{figure}

As illustrated in Figure \ref{fig:MicroArchitecture}, the matrix unit consists primarily of the Memory Loader, Scratchpad, Data Controller, and PE Array.

\textbf{Memory Loader} generates memory access requests and handles all data reads and writes for matrix computations. Within it, the Request Generator translates the tensor described by matrix multiplication instructions into memory and Scratchpad addresses, generating and issuing the corresponding requests. The Data Reorder module receives the returned data and reorder it as required, and writes it to the target storage.

\textbf{Scratchpad} temporarily stores matrix partitions to improve data reuse in the matrix unit. Accumulation results can remain resident in the Scratchpad, thereby reducing the need to write back high-precision results. A multi-bank Scratchpad further enables overlapping of data loading and computation tasks.

\textbf{Data Controller} supplies data to the computation array. Three Data Controllers are instantiated: two dedicated to source matrices A and B in matrix multiplication, and one dedicated to bias and accumulation matrix C.

\textbf{PE Array} combines outer-product and vector dot-product operations, where operands A and B are broadcast row-wise and column-wise across the array, respectively. Each PE performs inner-product operations and supports mixed-precision computing for TF32, BF16, FP16, INT8, and FP8 formats. Within each PE, multiplication results are aligned to a common exponent, truncated, and accumulated. The PE is organized into a six-stage pipeline to achieve a 2 GHz operating frequency with 14nm process node.

\vspace{-4pt}
\subsection{Configurable Matrix Extension}

\begin{table}[htbp]
\vspace{-5pt}
\caption{Configurable Architectural Parameters.}
\vspace{-15pt}
\begin{center}
\begin{tabular}{c c c}
\midrule
\textbf{Parameters}&\textbf{Meaning}&\textbf{Case Study} \\
\midrule
\textit{Freq} & Clock Frequency & 2.0 GHz \\
${M_{pe}}$ & Row of PE Array & 4 \\
${N_{pe}}$ & Column of PE Array & 4 \\
${K_{pe}}$ & PE Reduce Width & 512 Bits  \\
${M_{scp}}$ & Max Resident M in Scratchpad & 64 \\
${N_{scp}}$ & Max Resident N in Scratchpad & 64 \\
${K_{scp}}$ & Max Resident K in Scratchpad & 64 Bytes \\
\midrule
\multicolumn{2}{c}{{Data Bandwidth}}  & 48 GB/s  \\
\multicolumn{2}{c}{{Throughput (8-bit)}}  & 4 TOPS  \\
\midrule
\end{tabular}
\label{tab:configureable_micro_param}
\end{center}
\vspace{-5pt}
\end{table}

In order to support the generation of implementations with different compute capabilities, we provide a set of configurable microarchitectural parameters to tune the matrix unit for the compute requirements of diverse SoCs.

\cref{tab:configureable_micro_param} lists the configurable microarchitectural parameters of the matrix unit, with a case-study configuration aligned with the compute throughput and data bandwidth of Intel Xeon 8580 AMX.

The throughput of the PE array is determined by ${M_{pe}}$, ${N_{pe}}$, and ${K_{pe}}$. For an n-bit data format, the theoretical throughput is:

\vspace{-10pt}
\begin{equation}
\text{Throughput (n-bit)} = Freq \times M_{pe} \times N_{pe} \times (K_{pe}/n) \times 2
\end{equation}

The parameters ${M_{scp}}$, ${N_{scp}}$, and ${K_{scp}}$ define the scratchpad size. By adjusting the scratchpad capacity, the data-reuse level of the matrix unit can be adjusted to match different memory-bandwidth constraints. To avoid wasted compute, it is necessary to ensure that, under the output-stationary scheduling strategy, the compute time in the matrix-multiplication loop does not exceed the memory-access time:

\vspace{-10pt}
\begin{equation}\label{equation_config}
\frac{M_{scp} \times N_{scp} \times K_{scp}}{Freq \times M_{pe} \times N_{pe} \times K_{pe}}
\le
\frac{(M_{scp} + N_{scp}) \times K_{scp}}{DataBandwidth}
\end{equation}

Here, $DataBandwidth$ denotes the data-supply bandwidth of the lower-level memory hierarchy, determined by the cache structure, on-chip network and QoS, memory bandwidth, and other related system factors.

\subsection{Vector-Matrix Overlap}

\begin{figure}[t]
\vspace{-15pt}
\centering
\includegraphics[width=\linewidth]{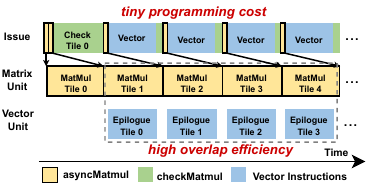}
\vspace{-20pt}
\caption{Vector and Matrix overlap.}
\label{fig:vec_mat_overlap}
\vspace{-9pt}
\end{figure}

The asynchronous matrix multiplication abstraction allows matrix tasks to be issued from the CPU pipeline without occupying it until completion, creating more opportunities for matrix–vector overlap. For the operators shown in Figure \ref{fig:ai_workload}, we adopt the programming method illustrated in Listing~\ref{code:code} to implement matrix–vector fused kernels, with the vector unit executing the prologue and epilogue while the matrix unit handling linear and convolution. Matrix multiplication and vector operations are orchestrated in a software pipeline at the granularity of matrix tiling, producing the fused execution behavior depicted in Figure \ref{fig:vec_mat_overlap}.

\vspace{-4pt}
\subsection{Low Design Overhead Integration}

The proposed matrix extension was integrated into the four open-source CPU RTL platforms listed in \cref{tab:integration_cost}, covering architectures from in-order single-issue to out-of-order six-issue designs. For Rocket, Shuttle, and BOOM, the integration reused the RoCC\cite{rocket} interface and was completed in a few days; for the XiangShan-Kunminghu processor, a new CSR interface was added, taking several weeks. Across all four processors, integrating the matrix extension required only 200–500 additional lines of RTL code, demonstrating the ease of integration of the proposed design.

\begin{table}[htbp]
\caption{Development Cost of Integration.}
\vspace{-15pt}
\begin{center}
\setlength{\tabcolsep}{1pt}
\begin{tabular}{ccccc}
\midrule
\textbf{CPU}&\textbf{Micro Architecture}&\textbf{Interface}&\textbf{Code*}&\textbf{Time} \\
\midrule
Rocket\cite{rocket} & In-order, 1-issue     & RoCC & 254 & 3 days  \\
Shuttle\cite{shuttle}           & In-order, 3-issue     & RoCC & 512 & 5 days  \\
BOOM\cite{boom}        & Out-of-order, 4-issue & RoCC & 301 & 3 days  \\
Xiangshan\cite{kunminghu}       & Out-of-order, 6-issue & CSR  & 361 & 3 weeks \\
\midrule
\end{tabular}
\footnotesize{*\textbf{Code} indicates the lines of integration-related RTL modifications.}
\label{tab:integration_cost}
\end{center}
\vspace{-10pt}
\end{table}

\vspace{-6pt}
\section{Evaluation}

\subsection{Methodology}

The experimental setup is shown in \cref{tab:exp_setup}, with Rocket, Shuttle, and BOOM evaluated on the Chipyard platform, and XiangShan-Kunminghu on the XiangShan platform. \cref{tab:exp_setup} also lists parameter configurations used to validate the scalability of the matrix extension. For reference, matrix extension configurations of three commercial CPUs are shown in the case-study column of \cref{tab:configureable_micro_param}.

\begin{table}[t]
\vspace{-5pt}
\caption{Experimental Setup and Configuration.}
\vspace{-15pt}
\begin{center}
\begin{tabular}{cc}
\midrule
\textbf{Parameter} & \textbf{Configuration} \\
\midrule
SoC Framework    & Chipyard\cite{chipyard} + Verilator + DRAMSim\cite{dramsim2} \\
CPU Core         & Rocket / Shuttle / BOOM / Kunminghu* \\
Vector Unit      & 512-bit RVV Saturn \cite{saturn} \\
PE Array Size    & 2×2 / 4×4 / 8×8 / 16×16 \\
PE Reduce Width  & 256 / 512 Bits \\
Clock Frequency  & 2.0 GHz \\
Data Bandwidth   & 8 / 16 / 32 / 48 / 64 GB/s \\
\midrule
\end{tabular}
\label{tab:exp_setup}
\footnotesize{*Kunminghu is evaluated on XiangShan\cite{kunminghu} SoC platform.}
\end{center}
\vspace{-6pt}
\end{table}

This work is compared against three commercial matrix extensions representative of prior work, as listed in \cref{sec:relatedwork}. All baselines are evaluated with 8-bit inference on ResNet-50 v1.5, BERT-base, and Llama3.2-1B, with Llama3.2-1B quantized using SmoothQuant-O1\cite{smoothquant} to maintain accuracy. Peak performance and per-core memory bandwidth are reported in \cref{tab:baseline_config}, 
with memory bandwidth measured using MLC\cite{intel_mlc} and STREAM\cite{stream}. IBM S1022 is evaluated with two threads to fully utilize the MMA unit\cite{hotchip_power10}. Intel AMX is tested with OneDNN v3.9\cite{onednn} and OpenVINO 2025.2.0\cite{openvino}; Arm SME with KleidiAI 1.14.0\cite{kleidiai} and ONNX Runtime (ORT) 1.21\cite{onnxruntime}; IBM MMA with OpenBLAS 0.3.27\cite{openblas} and ORT 1.16.3.

\begin{table}[t]
\vspace{-5pt}
\caption{Baseline configurations}
\vspace{-15pt}
\begin{center}
\setlength{\tabcolsep}{2.5pt}
\begin{tabular}{ccccc}
\midrule
\textbf{Platform} & \textbf{Framework} & \textbf{ISE} & \textbf{Bandwidth} & \textbf{INT8 Peak} \\
\midrule
IBM S1022 & ONNX Runtime\cite{onnxruntime} & MMA & 52.37 GB/s  & 2 TOPS   \\
Apple M4  & ONNX Runtime\cite{onnxruntime} & SME & 131.31 GB/s & 4 TOPS   \\
Xeon 8580 & OpenVINO\cite{openvino}     & AMX & 49.48 GB/s  & 4.6 TOPS \\
\midrule
\end{tabular}
\label{tab:baseline_config}
\end{center}
\vspace{-15pt}
\end{table}
\subsection{Integration Across Various CPU Platforms}\label{sec:gemm_test}

\begin{figure}[htbp]
\centering
\includegraphics[width=\linewidth]{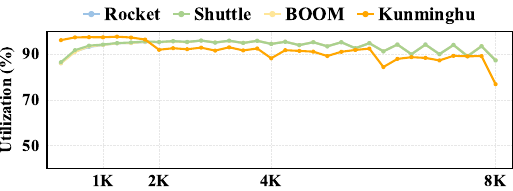}
    \vspace{-20pt}
\caption{GEMM Performance on Various CPU Platforms.}
    \vspace{-15pt}
\label{fig:diff_core_util}
\end{figure}

The matrix extensions on four CPU platforms including Rocket, Shuttle, BOOM, and Kunminghu are configured for a peak throughput of 2 TOPS, and GEMM workloads are evaluated with $M=512$, $N=512$, and $K$ ranging from 256 to 8192. As shown in Figure \ref{fig:diff_core_util}, all platforms achieve matrix unit utilization above 90\%. The results demonstrate the adaptability of our matrix extension across different CPU architectures.

\subsection{Matrix Unit Configurations at Various Scales}

\begin{figure}[t]
    \vspace{-15pt}
    \centering
    \includegraphics[width=\linewidth]{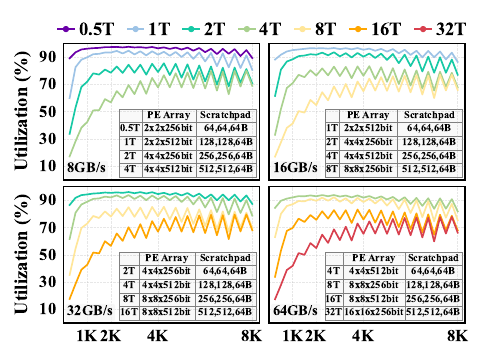}
    \vspace{-25pt}
    \caption{GEMM Performance under Different Config.}
    \vspace{-13pt}
    \label{fig:different_SOC_target}
\end{figure}

Matrix extension configurations are instantiated on the Shuttle CPU under four bandwidth settings, each targeting distinct peak compute capabilities.
As shown in Figure \ref{fig:different_SOC_target}, scratchpad sizes are configured according to Equation \ref{equation_config}. The GEMM workload follows \cref{sec:gemm_test}. Matrix unit utilization reaches approximately 80\% across all configurations. The proposed matrix extension adapts to both bandwidth-constrained embedded scenarios and high-performance CPUs, demonstrating strong scalability.

\subsection{Performance under Various Workloads}\label{sec:Performance_Comparison}

In this section, the matrix extension on Shuttle CPU is configured as in the case study of \cref{tab:configureable_micro_param} and paired with a 512-bit Saturn\cite{saturn} vector unit, to serve as a counterpart to the Xeon 8580 AMX extension. It is evaluated against the three commercial processors from \cref{tab:baseline_config} on four representative workloads.

\begin{figure}[htbp]
\vspace{-15pt}
\centering
\includegraphics[width=0.95\linewidth]{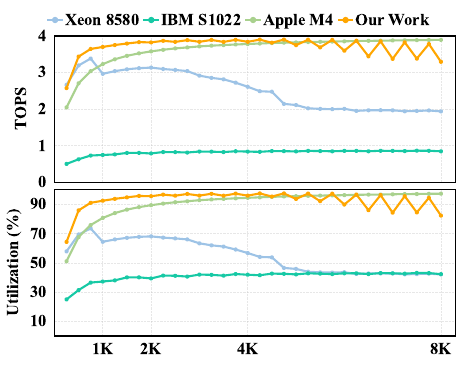}
\vspace{-15pt}
\caption{GEMM Performance vs. Existing Extensions.}
\label{fig:gemm_comp}
\vspace{-5pt}
\end{figure}

For \textbf{GEMM}, as shown in Figure \ref{fig:gemm_comp}, the proposed matrix extension outperforms the Xeon 8580 (OneDNN) and IBM S1022 (OpenBLAS), and approaches Apple M4 (KleidiAI). Performance fluctuations are caused by variations in DRAMSim memory throughput under different stride access patterns.

\begin{figure}[htbp]
\vspace{-5pt}
\centering
\includegraphics[width=\linewidth]{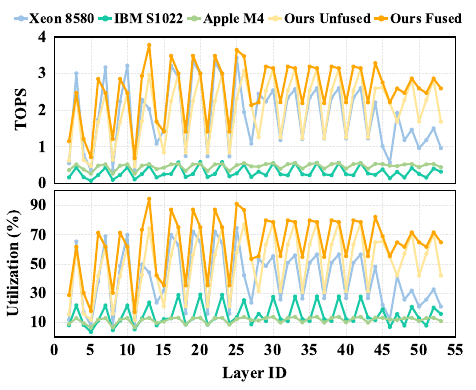}
\vspace{-25pt}
\caption{ResNet-50 Performance vs. Existing Extensions.}
\label{fig:resnet50_comp}
\vspace{-10pt}
\end{figure}

For \textbf{ResNet-50}, as shown in Figure \ref{fig:resnet50_comp}, AMX (OpenVINO) provides better operator support compared with MMA (ORT) and SME (ORT). Benefiting from 300 MiB L3 cache on Xeon 8580, all weights and activations can be buffered, enabling higher effective memory bandwidth. Currently, SME lacks support for convolution operators. Without vector-matrix fused kernels, the performance of our implementation is comparable to AMX. With fused kernels enabled, overall performance achieves speedups of 1.57×, 8.87×, and 5.04× compared to Xeon 8580, IBM S1022, and Apple M4, respectively.

\begin{figure}[t]
\vspace{-15pt}
\centering
\includegraphics[width=\linewidth]{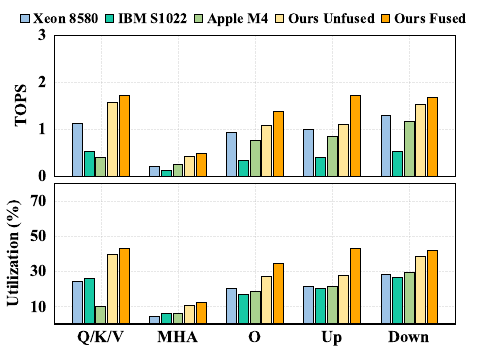}
\vspace{-20pt}
\caption{BERT Performance vs. Existing Extensions.}
\label{fig:bert_comp}
\vspace{-4pt}
\end{figure}

For \textbf{BERT}, as shown in Figure \ref{fig:bert_comp}, Transformer models dominated by small-scale matrix multiplications pose challenges for matrix unit utilization. Vendor extensions only achieve moderate utilization limited by small matrix multiplication sizes. With superior support for small-scale matrices, our implementation achieves speedups of 1.57×, 3.33×, and 2.11× compared to Xeon 8580 (OpenVINO), IBM S1022 (ORT), and Apple M4 (ORT), respectively.

\begin{figure}[t]
\centering
\vspace{-5pt}
\includegraphics[width=\linewidth]{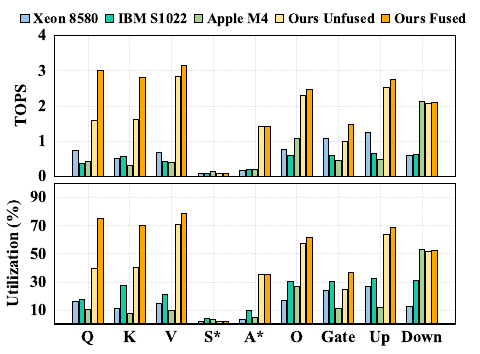}
\vspace{-25pt}
\caption{Llama3 Performance vs. Existing Extensions.}
\label{fig:llama_comp}
\vspace{-6pt}
\end{figure}

For \textbf{Llama3}, as shown in Figure \ref{fig:llama_comp}, the Llama3 model has a more complex Transformer structure than BERT and requires newer quantization methods such as SmoothQuant\cite{smoothquant} to control accuracy loss, which necessitates agile hardware–software co-optimization. Compared to BERT, Llama3 involves larger-scale matrix multiplications, enabling our matrix extension to achieve higher utilization. This also makes the execution time of vector operations comparable to matrix operations, yielding significant benefits from kernel fusion. Our matrix extension achieves speedups of 2.31×, 3.08×, and 3.16× compared to Xeon 8580 (OpenVINO), IBM S1022 (ORT), and Apple M4 (ORT), exceeding those observed for BERT. This result indicates that there is substantial room for hardware–software co-optimization of CPU matrix extensions on emerging workloads.

Our implementation exhibits relatively low performance on the Score (S*) operation due to Softmax dominance and limited vector throughput on Saturn. The significant gap between Gate and Up is due to the floating-point division in SiLU. On Saturn, vector division is implemented element-wise and is constrained by vector throughput. This highlights the optimization potential of current open-source RVV vector units.

\begin{table}[t]
\caption{Speed Up.}
\vspace{-15pt}
\begin{center}
\setlength{\tabcolsep}{3pt}
\begin{tabular}{c|ccc|ccc|ccc}
\midrule
&\multicolumn{3}{c}{\textbf{Xeon 8580}} & \multicolumn{3}{c}{\textbf{IBM S1022}} & \multicolumn{3}{c}{\textbf{Apple M4}} \\

& R & B & L & R & B & L & R & B & L \\
\midrule
Ours unfused &1.19 & 1.28 & 1.87 & 7.16 & 2.72 & 2.39 & 3.82 & 1.72 & 2.55 \\
Ours fused &1.57 & 1.57 & 2.31 & 8.87 & 3.33 & 3.08 & 5.04 & 2.11 & 3.16 \\
\midrule
\end{tabular}
\label{tab:sppedup}
\end{center}
\vspace{-5pt}
\footnotesize{R, B, and L denote ResNet-50, BERT-base, and Llama3.2-1B, respectively.}
\vspace{-10pt}
\end{table}

The speedups of our matrix extension over three commercial processor extensions are shown in \cref{tab:sppedup}. Current kernel implementations on commercial platforms do not fully leverage the matrix hardware. The substantial improvements observed in our fused implementation compared to unfused kernels suggest that the proposed programming model makes it easier to harness the full potential of matrix hardware.

\vspace{-5pt}
\subsection{Area and Power}

The matrix extension configurations from \cref{sec:Performance_Comparison} were synthesized in a 14nm process technology. The operating frequency reaches 2 GHz, and Table~\ref{tab:physical_evaluation} reports the area and power consumption.

\begin{table}[htbp]
\caption{Area and Power Evaluation (4TOPS@2GHz).}
\vspace{-15pt}
\begin{center}
\begin{tabular}{ccc}
\midrule
\textbf{} & \textbf{Area (mm\textsuperscript{2})} & \textbf{Power (W)} \\
\midrule
RAM & 0.164 & 0.784 \\
Logic & 0.367 & 0.722\\
\midrule
Total & 0.531 & 1.506\\
\midrule
\end{tabular}
\label{tab:physical_evaluation}
\vspace{-5pt}
\end{center}
\end{table}

\section{Conclusion}

Overall, this paper presents a unified and configurable CPU matrix extension architecture that facilitates agile cross-platform integration and efficient execution. The architecture has been deployed across multiple open-source CPU platforms, consistently delivering high performance under diverse configurations and workloads, thereby confirming its adaptability and scalability. This contribution offers the open-source community a practical solution for CPU matrix extensions. Future work will continue to explore the potential of CPU matrix extensions.

\section{Acknowledgement}
This work is supported by the ICT Innovation Grant E461100.

\clearpage

\bibliographystyle{ACM-Reference-Format}
\bibliography{sec_9_reference}

@misc{riscv_ime,
  title         = "Integrated Matrix Extension",
  author        = "Guido Araujo and Jose Moreira and Rafael Sene and Erich Focht",
  year          = 2025,
  howpublished={\url{https://github.com/riscv-admin/integrated-matrix-extension}},
}

@misc{riscv_vme,
  title         = "Vector-Matrix Extension",
  author        = "Greg Favor",
  year          = 2025,
  howpublished={\url{https://riscv.atlassian.net/wiki/spaces/VMEX/pages/554991628/Vector-Matrix+Extension+VME+-+PoW}},
}

@misc{riscv_ame,
  title         = "Attached Matrix Extension",
  author        = "Rafael Sene and Philipp Tomsich",
  year          = 2023,
  howpublished={\url{https://github.com/riscv-admin/attached-matrix-extension}},
}

@misc{intel_amx,
  title         = "Intel 64 and IA-32 Architectures Optimization Reference Manual",
  howpublished={\url{https://cdrdv2-public.intel.com/814201/355308-Optimization-Reference-Manual-049-Changes-Doc.pdf}},
  year          = 2024,
  author        = "Intel",
}

@misc{onnxruntime,
  title={ONNX Runtime},
  author={ONNX Runtime developers},
  year={2021},
  howpublished={\url{https://onnxruntime.ai/}},
}

@misc{openvino,
  title={Intel Distribution of OpenVINO Toolkit},
  author={OpenVINO developers},
  year={2025},
  howpublished={\url{https://github.com/openvinotoolkit/openvino}},
}

@misc{shuttle,
  title         = "Shuttle: A Rocket-based Superscalar In-order RISC-V Core",
  author        = "Jerry Zhao and Jennifer Zhou and Albert Ou and Abraham Gonzalez and Lux Zhang",
  year          = 2023,
  howpublished={\url{https://github.com/ucb-bar/shuttle/tree/master}},
}

@misc{intel_mlc,
  title         = "Intel Memory Latency Checker v3.12",
  author        = "Intel",
  year          = 2025,
  howpublished={\url{https://www.intel.com/content/www/us/en/developer/articles/tool/intelr-memory-latency-checker.html}},
}

@misc{kleidiai,
  title         = "KleidiAI",
  author        = "Arm",
  year          = 2025,
  howpublished={\url{https://gitlab.arm.com/kleidi/kleidiai}},
}

@misc{openblas,
  title         = "OpenBLAS",
  author        = "Zhang Xianyi, Martin Kroeker",
  year          = 2025,
  howpublished={\url{https://github.com/OpenMathLib/OpenBLAS}},
}

@misc{onednn,
author = "oneDNN Contributor",
year          = 2025,
title = {{oneAPI Deep Neural Network Library (oneDNN)}},
howpublished={\url{https://github.com/uxlfoundation/oneDNN}},
}

@INPROCEEDINGS{arm_sme,
  author={Remke, Stefan and Breuer, Alexander},
  booktitle={Workshops of the International Conference for High Performance Computing, Networking, Storage and Analysis(SC24-W)}, 
  title={Hello SME! Generating Fast Matrix Multiplication Kernels Using the Scalable Matrix Extension}, 
  year={2024},
  volume={},
  number={},
  pages={1443-1454},
  location = {Atlanta, GA, USA},
  publisher = {{IEEE} Press},
}

@ARTICLE{ibm_mma,
  title={A matrix math facility for Power ISA (TM) processors},
  author={Moreira, Jos{\'e} E and Barton, Kit and Battle, Steven and Bergner, Peter and Bertran, Ramon and Bhat, Puneeth and Caldeira, Pedro and Edelsohn, David and Fossum, Gordon and Frey, Brad and others},
  journal={arXiv preprint arXiv:2104.03142},
  year={2021},
}

@techreport{rocket,
    Author= {Asanović, Krste and Avizienis, Rimas and Bachrach, Jonathan and Beamer, Scott and Biancolin, David and Celio, Christopher and Cook, Henry and Dabbelt, Daniel and Hauser, John and Izraelevitz, Adam and Karandikar, Sagar and Keller, Ben and Kim, Donggyu and Koenig, John and Lee, Yunsup and Love, Eric and Maas, Martin and Magyar, Albert and Mao, Howard and Moreto, Miquel and Ou, Albert and Patterson, David A. and Richards, Brian and Schmidt, Colin and Twigg, Stephen and Vo, Huy and Waterman, Andrew},
    Title= {The Rocket Chip Generator},
    Year= {2016},
    Month= {Apr},
    Url= {http://www2.eecs.berkeley.edu/Pubs/TechRpts/2016/EECS-2016-17.html},
}

@inproceedings{boom,
  title={SonicBOOM: The 3rd Generation Berkeley Out-of-Order Machine},
  author={Zhao, Jerry and Korpan, Ben and Gonzalez, Abraham and Asanovic, Krste},
  booktitle={Fourth Workshop on Computer Architecture Research with RISC-V (CARRV)},
  year={2020},
  month={May}
}

@inproceedings{kunminghu,
author = {Xu, Yinan and Yu, Zihao and Tang, Dan and Chen, Guokai and Chen, Lu and Gou, Lingrui and Jin, Yue and Li, Qianruo and Li, Xin and Li, Zuojun and Lin, Jiawei and Liu, Tong and Liu, Zhigang and Tan, Jiazhan and Wang, Huaqiang and Wang, Huizhe and Wang, Kaifan and Zhang, Chuanqi and Zhang, Fawang and Zhang, Linjuan and Zhang, Zifei and Zhao, Yangyang and Zhou, Yaoyang and Zhou, Yike and Zou, Jiangrui and Cai, Ye and Huan, Dandan and Li, Zusong and Zhao, Jiye and Chen, Zihao and He, Wei and Quan, Qiyuan and Liu, Xingwu and Wang, Sa and Shi, Kan and Sun, Ninghui and Bao, Yungang},
title = {Towards Developing High Performance RISC-V Processors Using Agile Methodology},
year = {2023},
publisher = {{IEEE} Press},
booktitle = {Proceedings of the 55th Annual IEEE/ACM International Symposium on Microarchitecture(MICRO)},
pages = {1178–1199},
numpages = {22},
location = {Chicago, Illinois, USA},
}

@inproceedings{RiscvTPU,
author = {Gao, Peng and Liu, Yang and Sun, Haonan and Jiang, Jiang and Wang, Jun and Hong, Zonghui and Qu, Jiali},
title = {OASIS: A Commercial High Performance Terminal AI Processor Supporting RISC-V Tensor Extension Instructions},
year = {2025},
publisher = {{ACM}},
booktitle = {Proceedings of the 58th IEEE/ACM International Symposium on Microarchitecture (MICRO)},
pages = {1264–1283},
numpages = {20},
location = {New York, NY, USA},
}

@inproceedings{mlperf,
author = {Reddi, Vijay Janapa and Cheng, Christine and Kanter, David and Mattson, Peter and Schmuelling, Guenther and Wu, Carole-Jean and Anderson, Brian and Breughe, Maximilien and Charlebois, Mark and Chou, William and Chukka, Ramesh and Coleman, Cody and Davis, Sam and Deng, Pan and Diamos, Greg and Duke, Jared and Fick, Dave and Gardner, J. Scott and Hubara, Itay and Idgunji, Sachin and Jablin, Thomas B. and Jiao, Jeff and John, Tom St. and Kanwar, Pankaj and Lee, David and Liao, Jeffery and Lokhmotov, Anton and Massa, Francisco and Meng, Peng and Micikevicius, Paulius and Osborne, Colin and Pekhimenko, Gennady and Rajan, Arun Tejusve Raghunath and Sequeira, Dilip and Sirasao, Ashish and Sun, Fei and Tang, Hanlin and Thomson, Michael and Wei, Frank and Wu, Ephrem and Xu, Lingjie and Yamada, Koichi and Yu, Bing and Yuan, George and Zhong, Aaron and Zhang, Peizhao and Zhou, Yuchen},
title = {MLPerf inference benchmark},
year = {2020},
publisher = {IEEE Press},
booktitle = {Proceedings of the ACM/IEEE 47th Annual International Symposium on Computer Architecture(ISCA)},
pages = {446–459},
numpages = {14},
location = {Virtual Event},
}

@article{Mixed-precisionMatrixMultiplication,
title = {The cambrian explosion of mixed-precision matrix multiplication for quantized deep learning inference},
journal = {Future Generation Computer Systems},
year = {2026},
author = {Héctor Martínez and Adrián Castelló and Francisco D. Igual and Enrique S. Quintana-Ortí},
publisher = {Elsevier},
}

@inproceedings{resnet50,
  author       = {Kaiming He and
                  Xiangyu Zhang and
                  Shaoqing Ren and
                  Jian Sun},
  title        = {Deep Residual Learning for Image Recognition},
  booktitle    = {2016 {IEEE} Conference on Computer Vision and Pattern Recognition (CVPR)},
  location     = {Las Vegas, NV, USA},
  pages        = {770--778},
  publisher    = {{IEEE}},
  year         = {2016},
}

@inproceedings{BERT,
  author       = {Jacob Devlin and
                  Ming{-}Wei Chang and
                  Kenton Lee and
                  Kristina Toutanova},
  title        = {{BERT:} Pre-training of Deep Bidirectional Transformers for Language
                  Understanding},
  booktitle    = {Proceedings of the 2019 Conference of the North American Chapter of
                  the Association for Computational Linguistics: Human Language Technologies (NAACL-HLT)},
  pages        = {4171--4186},
  publisher    = {Association for Computational Linguistics},
  year         = {2019},
  location     = {Minneapolis, MN, USA}
}

@Misc{grattafiori2024llama,
  title={The llama 3 herd of models},
  author={Grattafiori, Aaron and Dubey, Abhimanyu and Jauhri, Abhinav and Pandey, Abhinav and Kadian, Abhishek and Al-Dahle, Ahmad and Letman, Aiesha and Mathur, Akhil and Schelten, Alan and Vaughan, Alex and others},
  journal={arXiv preprint arXiv:2407.21783},
  year={2024}
}

@Misc{xFormers2022,
  author =       {Benjamin Lefaudeux and Francisco Massa and Diana Liskovich and Wenhan Xiong and Vittorio Caggiano and Sean Naren and Min Xu and Jieru Hu and Marta Tintore and Susan Zhang and Patrick Labatut and Daniel Haziza and Luca Wehrstedt and Jeremy Reizenstein and Grigory Sizov},
  title =        {xFormers: A modular and hackable Transformer modelling library},
  howpublished = {\url{https://github.com/facebookresearch/xformers}},
  year =         {2022}
}

@article{flashattention2,
  title={Flashattention-2: Faster attention with better parallelism and work partitioning},
  author={Dao, Tri},
  journal={arXiv preprint arXiv:2307.08691},
  year={2023}
}

@inproceedings{flashattention,
author = {Dao, Tri and Fu, Daniel Y. and Ermon, Stefano and Rudra, Atri and R\'{e}, Christopher},
title = {FLASHATTENTION: fast and memory-efficient exact attention with IO-awareness},
year = {2022},
publisher = {Curran Associates Inc.},
booktitle = {Proceedings of the 36th International Conference on Neural Information Processing Systems(NIPS)},
articleno = {1189},
numpages = {16},
location = {New Orleans, LA, USA},
}

@article{chipyard,
author = {Amid, Alon and Biancolin, David and Gonzalez, Abraham and Grubb, Daniel and Karandikar, Sagar and Liew, Harrison and Magyar, Albert and Mao, Howard and Ou, Albert and Pemberton, Nathan and Rigge, Paul and Schmidt, Colin and Wright, John and Zhao, Jerry and Shao, Yakun Sophia and Asanovi\'{c}, Krste and Nikoli\'{c}, Borivoje},
title = {Chipyard: Integrated Design, Simulation, and Implementation Framework for Custom SoCs},
year = {2020},
issue_date = {July-Aug. 2020},
publisher = {{IEEE} Press},
volume = {40},
number = {4},
journal = {IEEE Micro},
pages = {10–21},
numpages = {12}
}

@inproceedings{hotchip_power10,
  author       = {William J. Starke and
                  Brian W. Thompto},
  title        = {IBM's {POWER10} Processor},
  booktitle    = {{IEEE} Hot Chips 32 Symposium, {HCS} 2020, Palo Alto, CA, USA, August
                  16-18, 2020},
  pages        = {1--43},
  publisher    = {{IEEE}},
  year         = {2020},
  location      = {Palo Alto, CA, USA}
}

@ARTICLE{stream,
  author = {John D. McCalpin},
  title = {Memory Bandwidth and Machine Balance in Current High Performance
	Computers},
  journal = {IEEE Computer Society Technical Committee on Computer Architecture
	(TCCA) Newsletter},
  year = {1995},
  month = dec,
}

@inproceedings{SmoothQuant,
  author       = {Guangxuan Xiao and
                  Ji Lin and
                  Micka{\"{e}}l Seznec and
                  Hao Wu and
                  Julien Demouth and
                  Song Han},
  title        = {SmoothQuant: Accurate and Efficient Post-Training Quantization for
                  Large Language Models},
  booktitle    = {International Conference on Machine Learning(ICML)},
  series       = {Proceedings of Machine Learning Research},
  volume       = {202},
  pages        = {38087--38099},
  publisher    = {{PMLR}},
  year         = {2023},
  location      = {Honolulu, Hawaii, {USA}}
 }

@ARTICLE{dramsim2,
  author={Rosenfeld, Paul and Cooper-Balis, Elliott and Jacob, Bruce},
  journal={IEEE Computer Architecture Letters}, 
  title={DRAMSim2: A Cycle Accurate Memory System Simulator}, 
  year={2011},
  volume={10},
  number={1},
  pages={16-19},
}

@techreport{saturn,
    Author = {Zhao, Jerry and Grubb, Daniel and Rusch, Miles and Wei, Tianrui and Anderson, Kevin and Nikolic, Borivoje and Asanović, Krste},
    Title = {The Saturn Microarchitecture Manual},
    Institution = {EECS Department, University of California, Berkeley},
    Year = {2024},
    URL = {http://www2.eecs.berkeley.edu/Pubs/TechRpts/2024/EECS-2024-215.html},
}

@inproceedings{ktransformers,
author = {Chen, Hongtao and Xie, Weiyu and Zhang, Boxin and Tang, Jingqi and Wang, Jiahao and Dong, Jianwei and Chen, Shaoyuan and Yuan, Ziwei and Lin, Chen and Qiu, Chengyu and Zhu, Yuening and Ou, Qingliang and Liao, Jiaqi and Chen, Xianglin and Ai, Zhiyuan and Wu, Yongwei and Zhang, Mingxing},
title = {KTransformers: Unleashing the Full Potential of CPU/GPU Hybrid Inference for MoE Models},
year = {2025},
publisher = {{ACM}},
booktitle = {Proceedings of the ACM SIGOPS 31st Symposium on Operating Systems Principles (SOSP)},
pages = {1014–1029},
numpages = {16},
location = {Lotte Hotel World, Seoul, Republic of Korea},
}

@INPROCEEDINGS{hpca2020_me,
  author={Qin, Eric and Samajdar, Ananda and Kwon, Hyoukjun and Nadella, Vineet and Srinivasan, Sudarshan and Das, Dipankar and Kaul, Bharat and Krishna, Tushar},
  booktitle={2020 IEEE International Symposium on High Performance Computer Architecture (HPCA)}, 
  title={SIGMA: A Sparse and Irregular GEMM Accelerator with Flexible Interconnects for DNN Training}, 
  year={2020},
  volume={},
  number={},
  pages={58-70},
  location = {San Diego, CA, USA},
  }

@INPROCEEDINGS{gemmini,
  author={Genc, Hasan and Kim, Seah and Amid, Alon and Haj-Ali, Ameer and Iyer, Vighnesh and Prakash, Pranav and Zhao, Jerry and Grubb, Daniel and Liew, Harrison and Mao, Howard and Ou, Albert and Schmidt, Colin and Steffl, Samuel and Wright, John and Stoica, Ion and Ragan-Kelley, Jonathan and Asanovic, Krste and Nikolic, Borivoje and Shao, Yakun Sophia},
  booktitle={2021 58th ACM/IEEE Design Automation Conference (DAC)}, 
  title={Gemmini: Enabling Systematic Deep-Learning Architecture Evaluation via Full-Stack Integration}, 
  year={2021},
  volume={},
  number={},
  pages={769-774},
  location = {San Francisco, CA, USA}
  }

@INPROCEEDINGS{asplos2026_Configurationwall,
  title={The Configuration Wall: Characterization and Elimination of Accelerator Configuration Overhead},
  author={Van Delm, Josse and Lydike, Anton and Dumoulin, Joren and Crols, Jonas and Yi, Xiaoling and Antonio, Ryan and Woodruff, Jackson and Grosser, Tobias and Verhelst, Marian},
  year={2025},
  note="To appear in Proceedings of ASPLOS 2026"
}

@ARTICLE{jssc2023_me,
  author={Houshmand, Pouya and Sarda, Giuseppe M. and Jain, Vikram and Ueyoshi, Kodai and Papistas, Ioannis A. and Shi, Man and Zheng, Qilin and Bhattacharjee, Debjyoti and Mallik, Arindam and Debacker, Peter and Verkest, Diederik and Verhelst, Marian},
  journal={IEEE Journal of Solid-State Circuits}, 
  title={DIANA: An End-to-End Hybrid DIgital and ANAlog Neural Network SoC for the Edge}, 
  year={2023},
  volume={58},
  number={1},
  pages={203-215},
  }

@ARTICLE{jssc2024_me,
  author={Conti, Francesco and Paulin, Gianna and Garofalo, Angelo and Rossi, Davide and Di Mauro, Alfio and Rutishauser, Georg and Ottavi, Gianmarco and Eggiman, Manuel and Okuhara, Hayate and Benini, Luca},
  journal={IEEE Journal of Solid-State Circuits}, 
  title={Marsellus: A Heterogeneous RISC-V AI-IoT End-Node SoC With 2–8 b DNN Acceleration and 30\%-Boost Adaptive Body Biasing}, 
  year={2024},
  volume={59},
  number={1},
  pages={128-142},
  }

@inproceedings{aspdac2025_me,
author = {Yi, Xiaoling and Antonio, Ryan and Dumoulin, Joren and Sun, Jiacong and Van Delm, Josse and Pereira Paim, Guilherme and Verhelst, Marian},
title = {OpenGeMM: A Highly-Efficient GeMM Accelerator Generator with Lightweight RISC-V Control and Tight Memory Coupling},
year = {2025},
publisher = {{ACM}},
booktitle = {Proceedings of the 30th Asia and South Pacific Design Automation Conference},
pages = {1055–1061},
numpages = {7},
location = {Tokyo, Japan},
}

\end{document}